\documentclass[a4paper,11pt]{article}
\pdfoutput=1 

\usepackage{jcappub}

\usepackage[T1]{fontenc}

\title{Bayesian Deep-stacking for High-energy Neutrino Searches}

\author[a]{I. Bartos,}
\author[b]{M. Ackermann,}
\author[b,c]{M. Kowalski}

\affiliation[a]{Department of Physics, University of Florida, PO Box 118440, Gainesville, FL 32611-8440, USA}
\affiliation[b]{Deutsches Elektronen Synchrotron DESY, Platanenallee 6, 15738 Zeuthen, Germany}
\affiliation[c]{Institut für Physik, Humboldt-Universität zu Berlin, 12489 Berlin, Germany}

\abstract{Following the discovery of the brightest high-energy neutrino sources in the sky, the further detection of fainter sources is more challenging. A natural solution is to combine fainter source candidates, and instead of individual detections, aim to identify and learn about the properties of a larger population. Due to the discreteness of high-energy neutrinos, they can be detected from distant very faint sources as well, making a statistical search benefit from the combination of a large number of distant sources, a called {\it deep-stacking}. Here we show that a Bayesian framework is well-suited to carry out such statistical probes, both in terms of detection and property reconstruction. After presenting an introductory explanation to the relevant Bayesian methodology, we demonstrate its utility in parameter reconstruction in a simplified case, and in delivering superior sensitivity compared to a maximum likelihood search in a realistic simulation.}

\begin{document}
\maketitle


\section{Introduction}

It has been more than a decade since IceCube identified a flux of high-energy neutrinos of cosmic origin \cite{2013Sci...342E...1I}. While the first neutrinos were detected as a quasi-diffuse flux without known origin, since then multiple individual neutrino sources have been identified, including the blazar TXS 0506+056 \cite{2018Sci...361.1378I,2018Sci...361..147I} and the active galaxy NGC 1068 \cite{2022Sci...378..538I}. A growing number of additional blazars and non-blazar active galactic nuclei have also been associated with detected neutrinos (e.g., \cite{2016NatPh..12..807K,2024PhRvL.132j1002N,2024PhRvD.110l3014K,2024arXiv240606684A}). Despite these identifications, the vast majority of astrophysical neutrinos still have no known origin, and additional source types may be necessary to explain the entire observed cosmic neutrino flux \cite{2021ApJ...921...45B}.

Due to the weakness of individual neutrino sources, previous identifications have typically required multi-messenger information. The first identified source, TXS 0506+056, was found by searching for electromagnetic counterparts temporally and directionally coincident with an energetic neutrino. Other searches considered limited candidate samples, such as bright gamma-ray emitters (e.g., \cite{2022Sci...378..538I}). Further sensitivity improvements were achieved by incorporating prior expectations of neutrino flux based on electromagnetic emission (e.g., \cite{2024PhRvD.110l3014K,2024arXiv240606684A}).

Further search sensitivity could be gained by using not just information of the source position, but also incorporating information on the expected neutrino flux, estimated from the candidates' electromagnetic emission, as a prior (e.g., \cite{2024PhRvD.110l3014K,2024arXiv240606684A}).

While the brightest neutrino sources have been identified, most sources are too distant and weak to be individually detected. A strategy to learn about these weaker sources is to combine information from them and target not individual but statistical detection. This strategy, typically called {\it stacking}, has been used extensively in neutrino searches (e.g., \cite{2020PhRvL.124e1103A,2024arXiv240606684A}). 

In the case of neutrino observations, a single high-energy neutrino can represent a detection due to the low background rate. Consequently, neutrino sources from virtually any distance can be observed, since there is always a non-zero chance that a single neutrino is detected. This is in contrast with most electromagnetic observations which have a threshold detectable flux, making distant sources impossible to observe.

The contribution of distant, weak sources to the overall detected high-energy flux is, therefore, significant. As discussed in \cite{2025arXiv250110213K}, adding sources out to a redshift of $z \sim 0.3$ can greatly increase search sensitivity, with diminishing returns for higher redshift. This corresponds to roughly $\sim 10-100$ blazars (assuming a local number density of $10^{-9}-10^{-8}$\,Mpc$^{-3}$), or $\sim10^{4}$ radio-loud AGN (assuming a local number density of $10^{-6}$\,Mpc$^{-3}$). Consequently, stacking searches optimally need to aim to incorporate a large number of source candidates over a broad distance scale. We refer to this strategy as {\it deep-stacking}. 

In this paper we present a Bayesian inference framework to carry out deep-stacking high-energy neutrino searches using source catalogs. In a deep-stacking search we expect the luminosity difference between the brightest and faintest sources to be substantial, making the effective utilization of source catalogs critical. Therefore, a key feature of a deep-stacking search is that expectations about the neutrino flux for each source needs to be taken into account. These expected fluxes are, however, highly uncertain. While we may know source distances precisely, we generally lack an accurate neutrino emission model that could convert known source properties to neutrino luminosity. This is the most straightforward in a Bayesian framework in which such expectations and their uncertainties can be readily incorporated in the analysis as priors. We demonstrate the benefit of a Bayesian framework in accounting for model uncertainties as nuisance parameters by comparing its accuracy in reconstructing a source parameter to a standard maximum likelihood approach.

Another important feature is that, since we aim to observe a population without the ability to resolve individual sources, scientific insight needs to be gained by understanding the properties of the population. This makes a hierarchical Bayesian framework particularly useful, since population properties, such as overall luminosity or redshift evolution, are straightforward to extract from observations.

Bayesian models have previously been used in astrophysical analyses, for example in gravitational wave population inference (e.g., \cite{2019MNRAS.486.1086M,2021ApJ...913L...7A,2017PhRvD..95j3010K,2017Natur.548..426F}). Similar models have also been previously proposed for neutrino analyses. \cite{2020PhRvD.101l3017C} implemented a Bayesian hierarchical model to place constraints on astrophysical neutrino populations. To limit computational cost and utilize publicly available data, this model built on the reconstructed flux and spectral index for astrophysical sources. \cite{2024arXiv240614268C} proposed a more general approach that directly incorporated the properties of each detected neutrino in a hierarchical Bayesian model, and demonstrated the applicability of such an approach to a population of sources. This latter work targets a similar direction as the present paper, with similar likelihood functions (c.f. Eq. \ref{eq:newbraun} and their Eq. 5). 

The aim of this paper is to facilitate the use of Bayesian deep-stacking methods in neutrino multi-messenger searches and parameter estimation. To achieve this, we first provide a clear conceptual introduction to Bayesian stacking, along with a detailed prescription for addressing key questions specific to deep-stacking analyses. We then perform a comprehensive simulation to demonstrate the advantages of Bayesian inference in a realistic scenario, explicitly comparing its performance to traditional frequentist approaches. Compared to previous hierarchical Bayesian analyses in neutrino astrophysics \cite{2020PhRvD.101l3017C,2024arXiv240614268C}, our primary novel contributions include explicitly addressing uncertainties in individual source luminosities and showcasing hierarchical Bayesian parameter estimation in a realistic context, directly contrasted with standard frequentist methods.

The paper is organized as follows. Section \ref{sec:method} introduces Bayesian stacking for high-energy neutrino searches, and gives prescriptions to the questions that can be addressed in this framework. Section \ref{sec:examples} presents results for a few demonstrative example inferences on simplified simulated sources.  Section \ref{sec:simulation} presents a detailed simulation and inference on a realistic source model. Finally, we conclude in Section \ref{sec:conclusion}. 

\section{Method}
\label{sec:method}

\subsection{Types of information}

We have the following types of parameters.

\paragraph{Neutrino Observation Parameters ($\mathbf{x}$).} These parameters are directly obtained from neutrino detection data. For 
detected neutrino event $i$, we denote these parameters as $x_i$, typically including the reconstructed neutrino energy $E_i$, the reconstructed sky location $\omega_i$, and the directional uncertainty $\sigma_i$. The neutrino observational data ensemble is then given by $\mathbf{x} \equiv \{x_1, x_2, \dots, x_{N_\nu}\}$. Alternative neutrino event properties or other proxies (e.g., energy estimators) can similarly be included depending on dataset specifics.

\paragraph{Fixed Source Parameters ($\xi_{\rm fix}$).} These are source properties assumed to be known or constrained independently from non-neutrino observations. Examples include source sky locations, luminosity distances, or host galaxy redshifts. These parameters are either assigned specific values or described by probability distributions derived from electromagnetic observations or theoretical models. Importantly, their values are not inferred from neutrino data and remain fixed throughout our inference.

\paragraph{Nuisance Parameters ($\xi_{\rm nui}$).} These represent unknown source properties that are not of primary interest in the analysis, but must be accounted for. Examples include individual source luminosities or intrinsic source spectral shape parameters if these are not the direct targets of the analysis. Since our primary analysis goals do not require the determination of these parameters, we marginalize over them, incorporating their uncertainties without explicitly reporting their inferred values.

\paragraph{Population (Hyper-)Parameters ($\xi_{\rm var}$).} These parameters characterize the neutrino source population rather than individual sources, and are explicitly targeted by the analysis. Examples include the average luminosity of the source population, intrinsic source rate densities, or parameters describing the population-level spectral distribution. The use of hyperparameters offers several advantages: (i) they influence multiple observations simultaneously, providing stronger statistical constraints; (ii) they reduce the complexity of the model, mitigating overfitting; and (iii) they yield insights into population-level properties rather than individual events.

Bayesian inference that includes hyperparameters is referred to as hierarchical Bayesian inference, which we focus on in the following.

\subsection{Bayesian inference}

As a first step in the Bayesian analysis, our aim is to compute the posterior probability densities of the unknowns $\xi_{\rm var}$  and $\xi_{\rm nui}$, i.e. $p(\xi_{\rm var},\xi_{\rm nui}|x_i,\xi_{\rm fix})$. Using Bayes' rule, we can write
\begin{equation}
p(\xi_{\rm var},\xi_{\rm nui}|\mathbf{x},\xi_{\rm fix}) = \frac{p(\mathbf{x}|\boldsymbol{\xi})\pi(\xi_{\rm var},\xi_{\rm nui}|\xi_{\rm fix})}{\pi(\mathbf{x}|\xi_{\rm fix})}
\label{eq:Bayes}
\end{equation}
where we use $\pi$ for priors, and we adopted $\boldsymbol{\xi}\equiv\{\xi_{\rm var},\xi_{\rm nui},\xi_{\rm fix}\}$ to simplify notation. 

\subsection{Priors}

The denominator in Eq. \ref{eq:Bayes} is the so-called {\it evidence}, i.e., the prior to observing $\mathbf{x}$. This is computed by marginalizing over the numerator of Eq. \ref{eq:Bayes}, i.e.:
\begin{equation}
\pi(\mathbf{x}|\xi_{\rm fix}) = \iint p(\mathbf{x}|\boldsymbol{\xi})\pi(\xi_{\rm var},\xi_{\rm nui}|\xi_{\rm fix})\,d\xi_{\rm var}d\xi_{\rm nui}.
\end{equation}
It is worth noting that this evidence typically does not need to be computed. For example, if the posterior is used for model comparison, then this cancels out when taking the ratio of two posteriors. As another example, if we are interested in the posterior probability density of $\xi_{\rm var}$ then the evidence will simply be a normalization factor.

The second term in the numerator of Eq. \ref{eq:Bayes} is the prior of $\xi_{\rm var}$ and $\xi_{\rm nui}$. It can be determined based on any prior information we assume on the distribution of these properties, or can be chosen as a noninformative (e.g., Jeffries) prior. For example, if the overall neutrino flux from our source catalog is one of the properties, one possible prior is a uniform distribution between 0 and the overall detected quasi-diffuse neutrino flux.

\subsection{Likelihood}

The term $p(\mathbf{x}|\boldsymbol{\xi})$ in Eq. \ref{eq:Bayes} is called the likelihood, and is the heart of the inference. To describe it it is worth specifying a few source parameters. Let $\hat{n}_j(\boldsymbol{\xi})$ be the {\it expected} number of detected neutrino events from source $j$, $\hat{n}_{\rm bg}(\boldsymbol{\xi})$ the expected number of detected background events, and $\hat{n}_{\rm tot}=\sum_j \hat{n}_j + \hat{n}_{\rm bg}$. Let $S_j(x_i|\boldsymbol{\xi})$ be the probability density of a detected astrophysical neutrino from source $j$ having $x_i$ reconstructed properties, and $B(x_i,\boldsymbol{\xi})$ the same for background neutrinos. Assuming $S_j$ and $B$ do not depend on uncertain source parameters we are probing with the analysis, the likelihood can then be written as (e.g., \cite{2024arXiv240614268C})
\begin{equation}
p(\mathbf{x}|\boldsymbol{\xi}) \propto e^{-\hat{n}_{\rm tot}}\prod_{i=1}^{n_\nu} \left[\sum_{j}^{N_{\rm s}}\hat{n}_j S_j(x_i|\boldsymbol{\xi}) + \hat{n}_{\rm bg}B(x_i,\boldsymbol{\xi})\right]
\label{eq:newbraun}
\end{equation}
where in the proportionality we omitted a factor that depends on the number of detected neutrinos but not on the source properties. Note that the factor $e^{-\hat{n}_{\rm tot}}$ appears here as the Poisson "$e^{-\lambda}$" term, while the product corresponds to the Poisson "$\lambda^k$" term.

The signal term $S_j(x_i|\boldsymbol{\xi})$ in Eq. \ref{eq:newbraun} can be written as a combination of probability densities from each source in the catalog:
\begin{equation}
S_j(x_i|\boldsymbol{\xi})=\frac{1}{2\pi \sigma_{i}^{2}}\exp\left(-\frac{|\omega_i - \omega_{{\rm s},j}|^2}{2\sigma_{i}^2}\right)P(E_i|\gamma_{\rm s})
\label{eq:likelihood}
\end{equation}
where $P(E_i|\gamma_{\rm s})$ is the probability density of measuring $E_i$ neutrino energy from a power-law distribution with index $\gamma_{\rm s}$ (assumed to be the same for each source), and $\omega_{{\rm s},j}$ is the true direction of source $j$. 

\subsection{Marginalization over nuisance parameters}

Once we obtain the posterior $p(\xi_{\rm var},\xi_{\rm nui}|\mathbf{x},\xi_{\rm fix})$, we need to marginalize over the nuisance parameters $\xi_{\rm nui}$ to obtain the posteriors of the properties we are interested in:
\begin{equation}
p(\xi_{\rm var}|\mathbf{x},\xi_{\rm fix})=\int p(\xi_{\rm var},\xi_{\rm nui}|\mathbf{x},\xi_{\rm fix})d\xi_{\rm nui}
\label{eq:marginalization}
\end{equation}

\subsection{Detection through model comparison}
\label{sec:model_comparison}
In addition to reconstructing a particular parameter, a relevant question can be whether the detected neutrinos can be used to claim the detection of a source population. This question is equivalent to asking whether the data supports at least some of the source luminosities to be greater than zero, compared to that null hypothesis that all luminosities are zero.  That is, we are interested in the odds ratio
\begin{equation}
\mathcal{O} = \frac{p(L_{\rm tot}>0|\mathbf{x},\xi_{\rm fix})}{p(L_{\rm tot}=0|\mathbf{x},\xi_{\rm fix})} = \frac{p(\mathbf{x}|L_{\rm tot}>0,\boldsymbol{\xi})}{p(\mathbf{x}|L_{\rm tot}=0,\boldsymbol{\xi})}\cdot\frac{\pi(L_{\rm tot}>0)}{\pi(L_{\rm tot}=0)}
\label{eq:oddsratio}
\end{equation}
where $L_{\rm tot}$ is the total luminosity of the investigated source population. 


A common method to interpret the odds ratio is to consider threshold values over which there is a significant preference for one model over the other. While the odds ratio naturally lends itself to a Bayesian interpretation--namely, directly providing the probability that one model is favored given the observed data--such an interpretation cannot be directly compared to widely-used frequentist significance metrics, in particular the false alarm rate, that are standard in neutrino and multi-messenger astrophysics. To facilitate this comparison and ensure that our results align with established frequentist practices, we instead use the odds ratio as a test statistic and derive its distribution under the background-only scenario. This allows our results to be interpreted within a frequentist framework, thus enabling a direct comparison with existing frequentist significance estimates in the literature.

\section{Demonstration through a simplified example}
\label{sec:examples}

To demonstrate the usage of the above framework, we consider a simplified neutrino detection model in which the demonstration of the calculation is straightforward. Specifically, we consider detected neutrinos whose only property is their association with a source within the catalog. Therefore, for each source, our observed quantity will be the number of associated neutrinos, with no further differentiator. We further assume that the background rate is the same for all sources. For this example we examine the reconstruction of the average source luminosity $L_0$.

\subsection{Monte Carlo simulation}

We evaluated each example below using Monte Carlo simulations to demonstrate the application the Bayesian inference framework described above. We considered $N_{\rm s}=1000$ sources whose location was randomly drawn from a uniform volumetric distribution within maximum distance $d_{\rm max}$. 

For a given realization, for each source, we drew a luminosity $L$ from a log-normal distribution with parameters $\mu_{\rm L}\equiv \ln(L_0)$ and $\sigma_{\rm L}$, which are the expected value and mean of $\ln(L)$. We used a representative value $\sigma_{\rm L}=0.3$, while multiple $L_0$ values were evaluated to examine the results dependence on the sources' signal-to-noise ratio. For source $j$ with luminosity $L_{j}$ and distance $d_j$, the expected number of neutrinos was computed as 
\begin{equation}
\hat{n}_{j} \propto L_{j} / (4 \pi d_{j}^{2}).
\end{equation}

We then drew numbers of detected neutrinos $n_j$ drawn from a Poisson distribution with mean $\hat{n}_{j}$. This included background detections as well. We assumed the background to be $\hat{n}_{\rm bg} B/ N_{\rm s} = 1$, i.e. that during the observation time each source will have, on average, 1 detected background neutrino.

\subsection{Average luminosity}
\label{sec:averageluminosity}

In this example, our variable of interest was $L_0$, therefore $\xi_{\rm var}\equiv L_0$. Assuming that each source's luminosity is randomly drawn from a distribution centered around $L_0$, the individual luminosity $L_j$ of source $j$ is also an unknown variable. However, since observations are typically not sensitive enough to know these individual luminosities, these will be nuisance parameters: $\xi_{\rm nui} \equiv \{L_j\}$. Finally, we assume that we know the luminosity distance $d_j$ for each source. Therefore, these source distances will be our fixed parameters: $\xi_{\rm fix} \equiv \{d_j\}$.

Let us start with the likelihood in Eq. \ref{eq:likelihood}. Here, $S_j(x_i|\boldsymbol{\xi})$ and will be 1 if $x_i$ is associated with source $j$, otherwise it will be 0. The background term $B(x_i,\boldsymbol{\xi})$ will be constant. To simplify our notation, we will use $\hat{n}_{j}'\equiv \hat{n}_j + \hat{n}_{\rm bg}B/N_{\rm s}$, i.e. we fold in the background expectation into our source-specific term. With this, we can write Eq. \ref{eq:likelihood} as
\begin{equation}
p(\mathbf{x}|\boldsymbol{\xi}) \propto e^{-\hat{n}_{\rm tot}} \prod_{j}^{N_{\rm s}} (\hat{n}_{j}')^{n_j} = \prod_{j}^{N_{\rm s}} e^{-\hat{n}_{j}'}(\hat{n}_{j}')^{n_j}= \prod_{j}^{N_{\rm s}} \mbox{Poiss}(n_j;\hat{n}_{j}') \, n_{j}!
\label{eq:simplelikelihood}
\end{equation}
Since the observed value $n_j$ does not depend on $\boldsymbol{\xi}$, we can multiply the likelihood with $n_{j}!^{-1}$ without changing the shape of the distribution. Therefore, we find that our likelihood is proportional to the product of the Poisson probabilities of observing $n_{j}$ events for an expected value of $\hat{n}_{j}'$. 

Once we obtained $n_j'$ and $\hat{n}_{j}$, we computed the likelihood using Eq. \ref{eq:simplelikelihood}. For the prior $\pi(\xi_{\rm var},\xi_{\rm nui}|\xi_{\rm fix})$, we assumed that this prior can be written as  $\pi(\xi_{\rm var})\pi(\xi_{\rm nui})$. We adopted a non-informative uniform distribution for the luminosity: $\pi(\xi_{\rm var}) = 2\exp(L_0+2\sigma_{\rm L})$. For $\pi(\xi_{\rm nui})$ we considered that we know the underlying log-normal distribution and adopted it as our prior. The evidence in the denominator of Eq. \ref{eq:Bayes} does not appear in our calculation and therefore we did not need to compute it.

We then computed the marginalized probability density (see Eq. \ref{eq:marginalization})
\begin{eqnarray}
p(L_0|\{n_{j}\},\{d_j\}) \propto \int \prod_{j}^{N_{\rm s}} \mbox{Poiss}(n_j;\hat{n}_{j}') p(L_j|L_0,\sigma_{\rm L}) \, \mathrm{d}L_j
\end{eqnarray}
While this is an $N_{\rm s}$ dimensional integral, these integrals can be independently computed\footnote{The integrals are independent if and only if sources do not overlap, i.e. if for each neutrino we can unambiguously determine which source they may correspond if they are from a source within the catalog. This is a reasonable assumption if the number of sources is much smaller than the total sky area divided by the number of sources times the typical size of a point spread function of a neutrino.} and then multiplied to get a final probability density for $L_0$, making the problem computationally tractable even for large number of sources. 

\begin{figure}[h]
    \centering
    \includegraphics[width=0.8\textwidth]{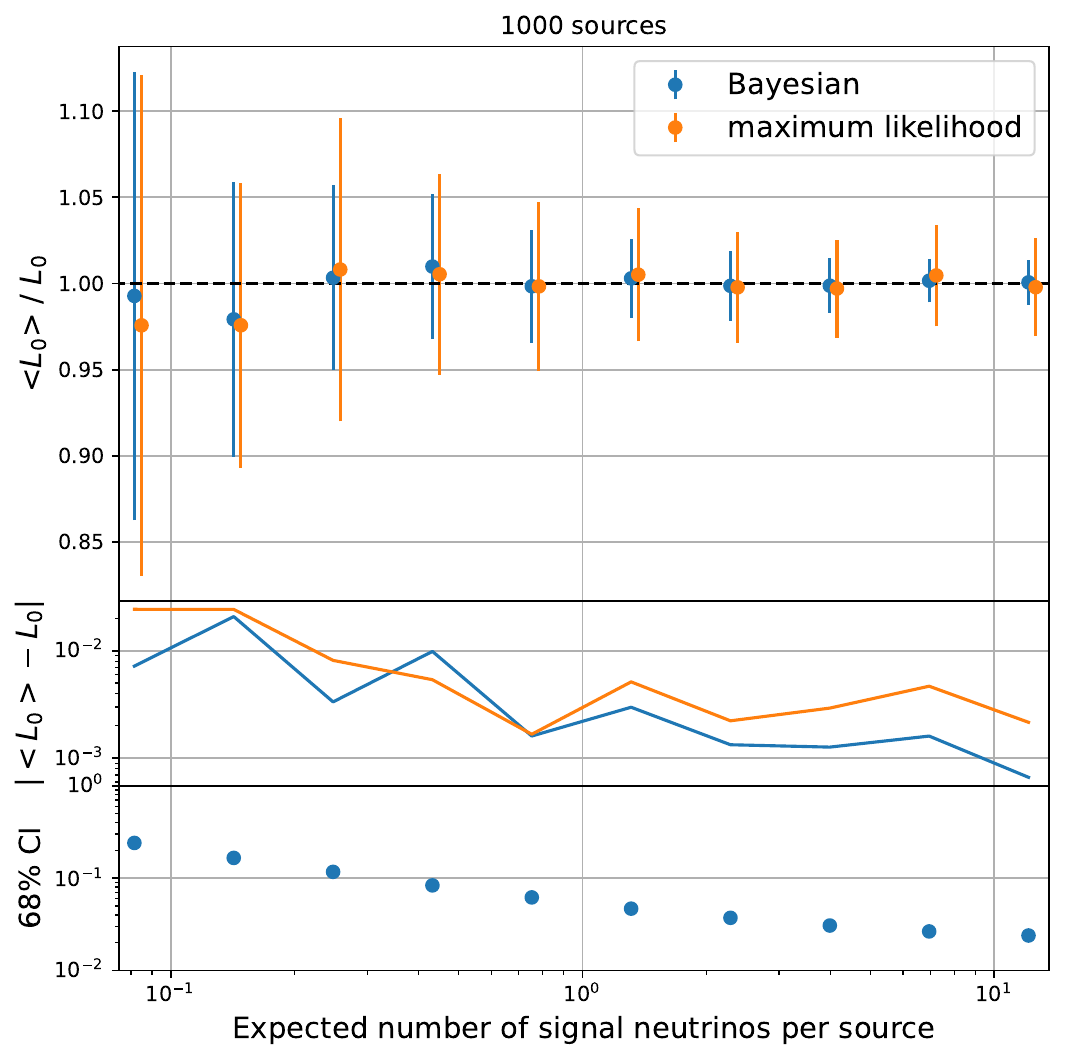}
    \caption{Comparison of the reconstructed average source luminosity, $\langle L_0 \rangle$, to the true luminosity, $L_0$, as a function of the expected number of detected astrophysical neutrinos per source. (top) reconstructed luminosity normalized by the true luminosity (the dashed line represents perfect reconstruction). (middle) absolute deviation of the reconstructed luminosity from the true value. (bottom) size of the 68\% credible interval (CI) for the Bayesian method. The analysis assumes 1000 cataloged neutrino sources and an average background level of 1 neutrino event per source. Results from Bayesian inference (blue) and maximum likelihood estimation (orange) methods are shown for comparison.}
    \label{fig:Lavg}
\end{figure}

Finally, we are interested in the expected value of $L_0$, which will be
\begin{equation}
\langle L_0 \rangle = \int L_0 \, p(L_0|\{n_{j}\},\{d_j\}) dL_0.
\end{equation}
Our results are shown in Fig. \ref{fig:Lavg} for a range of $L_0$ values, quantified for simplicity as the expected number of detected signal neutrinos per source. For comparison we also show estimates derived using a standard maximum likelihood method. For each point in Fig. \ref{fig:Lavg}, we carried out 100 independent simulations. The main takeaways of these results are the following: (1) We find that the average estimated $L_0$ over these simulations falls near the simulated value, showing that the method is unbiased. (2) As expected, the variance around this value, i.e. the typical deviation for individual simulations, gets gradually smaller as $L_0$ increases, i.e. the expected error of the reconstruction decreases. (3) We find that the Bayesian method generally performs better as the maximum likelihood method, as measured by the expected deviation from the true $L_0$ value (i.e., the error bars in Fig. \ref{fig:Lavg} (top) are smaller). (4) With the Bayesian method we also derive the 68\% credible interval for the $L_0$ reconstruction for each simulation, and show the average credible interval for each $L_0$ value in Fig. \ref{fig:Lavg} (bottom). This effectively characterizes the accuracy of the results, as "measured" by the Bayesian algorithm. We see that this interval decreases with increasing $L_0$.

\section{Demonstration through a realistic simulation}
\label{sec:simulation}
We created a realistic simulation to demonstrate how detection can be carried out using Bayesian deep-stacking, and to evaluate its sensitivity compared to standard frequentiest inference. We generated a realistic source population and conditions that are similar to what can be expected in current-generation neutrino telescopes. 

\subsection{Simulation setup}

Our simulation implements a version of a typical source-stacking analysis commonly applied in neutrino astronomy (e.g., \cite{Aartsen:2016lir},\cite{IceCube:2021pgw}) that, for simplicity, considers an idealized neutrino detector. In this framework, a sample of neutrino sources can be defined based on a few parameters: the luminosity and density of the population at $z=0$; an index $k$ that defines the redshift evolution factor $(1+z)^k$ of the sources within the region of interest $z\leq z_{max}=1$; a flag that selects pure density evolution (PDE) or pure luminosity evolution (PLE). 

The simulation also requires a set of parameters that define the observation: the fraction of the sky that is observable; the observation duration $t_{\mathrm{obs}}$\,; the spectral index $\gamma$ of the neutrino flux from this population; energy range, effective area and angular resolution of a detector that translates neutrino fluxes from sources into event counts. 

The background is modeled by two power-law components, a soft component reflecting the atmospheric background, and a hard component reflecting the astrophysical background from sources that are not part of the sample tested in the stacking analysis. The assumed effective area of the detector is defined as $A_{\mathrm{eff}} = A_{0} (E/E_{0})^{\delta}$, independent of the observation direction within the observable fraction of the sky, with $\delta=0.7$ set as default. This value approximates the increase of the effective area in neutrino detectors throughout the TeV energy range, due to the rise in the neutrino cross-sections and other effects such as an increased range of muons at higher energies.

The first step in the simulation chain is to generate a set of sources below a maximum redshift $z_{\mathrm{max}}$, where individual source fluxes $S_{j}$ are randomly sampled from the number density distribution $dN/dS$, which is defined by local source density $\rho_0$, luminosity $L_0$ and redshift evolution parameter $k$. Only sources in the observable fraction $f_{\mathrm{obs}}$ of the sky are used for the next steps. The generated sources define the population for the stacking analysis, and the expected number of neutrino events from these sources in a detector parametrized by ($A_{\mathrm{eff}}, t_{\mathrm{obs}}$) defines the set of optimal source weights $w^{(j)}_{\mathrm{opt}} \equiv \hat{n}_{j}$ (see below in Eq. \ref{eq:sample_stack}) used in the stacking analysis.

In the second step, observation pseudo-data (denoted as ``events'' below) are created for the source sample defined in step one. For each source, a number of events are generated from a Poisson distribution with mean $\hat{n}_{j}$. Each event is assigned an energy within the energy range $[E_{\mathrm{min}},E_{\mathrm{max}}]$, and a direction according to the specified angular resolution\footnote{The point spread function is modeled as a bivariate gaussian with $\sigma_1=\sigma_2=\Psi$} $\psi$. The energy spectrum of the events is sampled from a power law with index $\gamma$ 
that is common for all sources in the sample. 

Background events are added for both the hard and the soft background components, with directions drawn from an isotropic distribution within the observable fraction of the sky and energies sampled from power laws with indices $\alpha_{\mathrm{hard}},\alpha_{\mathrm{soft}}$ for the hard and soft components, respectively. 

\subsection{Frequentist likelihood analysis}
\label{sec:likelihood_analysis}

We perform a sequence of unbinned maximum-likelihood analyses on our simulated data sets to test the performance of a typical stacking analysis (and later compare to Bayesian deep-stacking). Only the expected numbers of events from the sources are fitted, while their spectral index $\gamma$ is fixed to the value used in the simulation. 

The most common approach in neutrino astronomy for a source stacking analysis, which we denote as \emph{sample normalization} uses the likelihood from Eq. \ref{eq:newbraun},
but instead of fitting the individual parameters $\hat{n}_{j}$ for each source, a single global normalization factor $\hat{n}_{\mathrm{sig}}$ for the entire sample is fitted, assuming that the relative expected contribution of each source in the sample is perfectly known from a-priori information. The likelihood function is then given by

\begin{equation}
    p(\mathbf{x}|\boldsymbol{\xi}) \propto e^{-(\hat{n}_{\mathrm{sig}}^{*} + \hat{n}_{\rm bg})}
    \prod_{i=1}^{n_\nu} \left[ 
        \hat{n}_{\mathrm{sig}}^{*} \,
         \frac{
        \sum_{j}^{N_{\rm s}} w^{(j)} S_j(x_i|\boldsymbol{\xi}) }{ \sum_{j}^{N_{\rm s}} w^{(j)} } 
        + \hat{n}_{\rm bg}  B(x_i,\boldsymbol{\xi})
     \right]
\label{eq:sample_stack}
\end{equation}

where the star in $\hat{n}_{\mathrm{sig}}^{*}$ indicates that this parameter is a fit parameter (to distinguish it from $\hat{n}_{\mathrm{sig}}$, which is the expected number of signal events from the source in the simulated population).
Of the signal PDF $S_j$ defined in Eq. \ref{eq:likelihood}, only the spatial term is used. Since it is computationally advantageous, and since the energy resolution of neutrino telescopes for high-energy neutrinos is typically poor, we split the events into $N_E$ energy bins instead of implementing an energy term in the likelihood. The likelihood is calculated separately for each energy bin, and the total likelihood is the product of the likelihoods in each energy bin:

\begin{equation}
    p(\mathbf{x}|\boldsymbol{\xi}) = \prod_{k=1}^{N_E} p_{k}(\mathbf{x}|\boldsymbol{\xi})
\end{equation}

The parameters $\hat{n}_{\mathrm{sig}}^{(k)} = f_{k}\, \hat{n}_{\mathrm{sig}}$ and $\hat{n}_{\mathrm{bg}}^{(k)} = g_{k}\, \hat{n}_{\mathrm{bg}}$ in the $p_{k}(\mathbf{x}|\boldsymbol{\xi})$ are the expected number of signal and background events in each energy bin. 
The $f_{k}$($g_{k}$) are the fraction of the total number of signal (background) events in energy bin $k$ and depend only on the assumed spectrum for the sources and the background. 
The spectral index is not fitted, therefore, the $f_{k}$ values are simple constants and independent of the fitted hypothesis.

The most important shortcoming of the sample normalization approach is its implicit assumption that the expected relative contribution of each source in the sample is known exactly. 
In practice, the expected neutrino flux from a source is estimated from a model that relates multi-wavelength observations, such as X-rays (e.g., \cite{2024PhRvD.110l3014K}), to the expected neutrino luminosity (e.g., $L_{\gamma} \propto L_{\nu}$). Uncertainties in this estimate, as well as source-by-source variations, are not taken into account. 

To test the robustness of the sample normalization approach, we performed a series of likelihood ratio tests with this approach varying the weights $w^{(j)}$ in the likelihood. 
In the ideal scenario we choose $w^{(j)} = w^{(j)}_{\mathrm{opt}} = \hat{n}_{j}$. To describe more realistic scenarios we set the $w^{(j)} = w^{(j)}_{\mathrm{opt}} \times 10^{r_j}$, where $r_j$ is drawn from a normal distribution with width $\sigma_S = \{0.3,0.5,1.0\}$ corresponding to factors of approximately 2, 3, and 10 variations of individual source luminosities with respect to the assumed correlation. 

\subsection{Frequentist likelihood analysis with source normalization}

An alternative maximum likelihood approach tailored to such realistic scenarios introduces individual parameters for each source as nuissance parameters with priors derived from the estimated scale of source-by-source variations. We denote this as \emph{source normalization} approach. The corresponding likelihood function is extended with respect to its form in Eq. \ref{eq:newbraun}, to
\begin{equation}
    q(\mathbf{x}|\boldsymbol{\xi}) \propto p(\mathbf{x}|\boldsymbol{\xi}) \, 
    \prod_{j}^{N_{\rm s}} \frac{1}{ \hat{n}_j^{*} \sigma_S}
    \exp\left(-\frac{(\log_{10} \hat{n}_j^{*} - \log_{10} \hat{n}_{j})^2}{2\sigma_S^2}\right)
\end{equation}

The likelihood is maximized with respect to the $N_{\rm s}$ parameters $\hat{n}_{j}^*$, which can become quite numerous in deep-stacking scenarios, and become challenging computationally. {\bf To match this challenge, the likelihood calculations for this manuscript have been optimized using the {\it PyTorch}\footnote{\url{https://pytorch.org
}} package that provides analytic gradients for the likelihood minimization and supports multi-core architectures as well as calculations of the likelihoods on specialized hardware such as GPUs.}

For either approach we calculate the test statistic as $\mathrm{TS}= -2 \log \left( p(\mathbf{x}|\boldsymbol{\xi}) / p(\mathbf{x}|\boldsymbol{0}) \right)$, where $\ p(\mathbf{x}|\boldsymbol{0})$ represents the null hypothesis $n_{j}^{*} \equiv 0$. The TS is calculated for the pseudo-data sets containing simulated neutrinos from the source population, as well as for background-only pseudo-data sets. The performance of each approach in each scenario is then compared on the basis of the expected p-value from this test. This p-value is calculated as the fraction of TS values from the background-only pseudo-data sets that are larger than the median TS value of the pseudo-data sets containing signal events. If the number of simulated datasets is not large enough to determine the p-value directly from the distributions, an extrapolation of the background-only TS distribution is used to estimate the p-value. For the extrapolation we use a $\chi^2$-distribution with one degree of freedom when applicable (sample normalization approach, source normalization approach with optimal weights), 
and a simple exponential function fitted to the tails of the distribution otherwise.

\subsection{Bayesian analysis}
For the Bayesian deep-stacking approach we used the odds ratio defined in Eq. \ref{eq:oddsratio} as a test statistic to compare to the frequentist likelihood maximization approaches and determine the performance of this method distinguishing the signal from the background-only sample in either scenario. A uniform prior was used for  $0 \leq \sum_j \hat{n}_j^{*} / \hat{n}_{\rm sig} \leq 2$, while a log-normal shape was used for the individual priors for $\hat{n}_j$, characterizing the expected source-by-source variations. Marginalizing over the parameters $\hat{n}_j^{*}$, the odds ratio becomes

\begin{equation}
    \mathcal{O} =  \frac{1}{p(\mathbf{x}|\boldsymbol{0},\xi_{\rm fix}) \pi(\boldsymbol{0})}
    \iint_{\hat{n}_j^{*}} 
    p(\mathbf{x}|\hat{n}_j^{*},\xi_{\rm fix})
    \pi(\hat{n}_j^{*})\;\;,\mathrm{ with}
\end{equation}

\begin{equation}
    \pi(\hat{n}_j^{*}) = \frac{1}{ \sqrt{2\pi} \hat{n}_j^{*} \sigma_S}
    \exp\left(-\frac{(\log_{10} \hat{n}_j^{*} - \log_{10} \hat{n}_{j})^2}{2\sigma_S^2}\right)\;\;, \mathrm{and} \;\;\pi(\boldsymbol{0})=1
\label{eq:bayes_prior}
\end{equation}

The odds ratio was calculated for the pseudo-data sets containing simulated neutrinos from the source population, as well as for background-only pseudo-data sets. As in the case for the frequentist likelihood ratio tests, we computed various scenarios of scattering between the injected source flux and the $\hat{n}_{j}$ used in the analysis model using the same set of widths, i.e., $\sigma_S = (0,0.3,0.5,1.0)$  

Using this odds ratio as a test statistic, the performance of the Bayesian deep-stacking approach for rejecting the null hypothesis was compared to the frequentist likelihood maximization approaches on the basis of the expected p-values.

For large numbers of sources, solving the $N_{S}$-dimensional integral in the odds ratio is generally computationally expensive. However, in the use case considered here, the signal PDF in the likelihood drops quickly with increasing distance from the source. 
Each event will only contribute to the signal PDF of a few sources, and the signal PDF of sources that are not close to the event direction will be negligible. 
This allowed us to reduce the dimensionality of the integral by only considering the source that is the closest to the event direction. This is justified in the case that the average distance between sources is much larger than the assumed angular resolution of our hypothetical detector and reduces the calculation of the odds ratio to calculating a series of one-dimensional integrals, {\bf which we perform  using simple Gaussian quadrature (QUADPACK~\cite{piessens1983quadpack}). 

Therefore our simulation code is currently limited to the conditions described above and does not scale favorably to even larger numbers of sources, when the average distance between sources approaches the angular resolution of the instrument. Multi-dimensional integrations are unavoidable in this case, and more advanced integration techniques, such as Monte Carlo integration, need to be evaluated and implemented to efficiently calculate the odds ratio in this case.}

\subsection{Results}

We carried out simulations assuming a local density of $10^{-6}$ Mpc$^{-3}$, which we sample up to a maximum redshift of z=0.1. We assume no redshift evolution ($k=0$), and the same neutrino luminosity of $10^{43}$\,erg\,s$^{-1}$ for all sources. The sources were assumed to emit neutrinos with a power law spectrum with an index of $\gamma$ = -2.5, motivated by the spectrum observed for the diffuse neutrino flux~\cite{IceCube:2020acn}. The observation time was set to 10 years, and the instantaneously observable fraction of the sky was $f_{\mathrm{obs}}=0.33$, typical values for today's high-energy neutrino telescopes, such as IceCube~\cite{2017JInst..12P3012A}or KM3NeT~\cite{KM3Net:2016zxf}. Events were recorded in the energy range between $\log_{10} (E/{\rm GeV})=3.5$ and $\log_{10} (E/{\rm GeV})=6.5$ and split into three bands, each spanning one decade in energy. The effective area parameter that relates source luminosity to expected number of events was set to $A_0=1.5\,\mathrm{m}^2$ at 10\,TeV. The simulated background was normalized to correspond approximately to the number of IceCube track events above 3\,TeV in~\cite{Abbasi:2021qfz}. This amounts to a total atmospheric background event density of $6.4 \times 10^3\;$sr$^{-1}$ simulated as a power law spectrum with index $-3.7$, and a total astrophysical background event density of $1.6 \times 10^2\;$sr$^{-1}$ simulated as a power law spectrum with index $-2.5$. 

Our setup yielded a total of $\approx$120 sources in the sample that we included in our various stacking analyses. The brightest source in each realization of the sample, if analyzed individually, only yields a median detection significance of about $2\sigma$, so that the population would remain entirely undetected in a large fraction of the simulations. 

\begin{figure}
    \centering
    \includegraphics[width=0.99\textwidth, trim=0 27 0 0, clip]{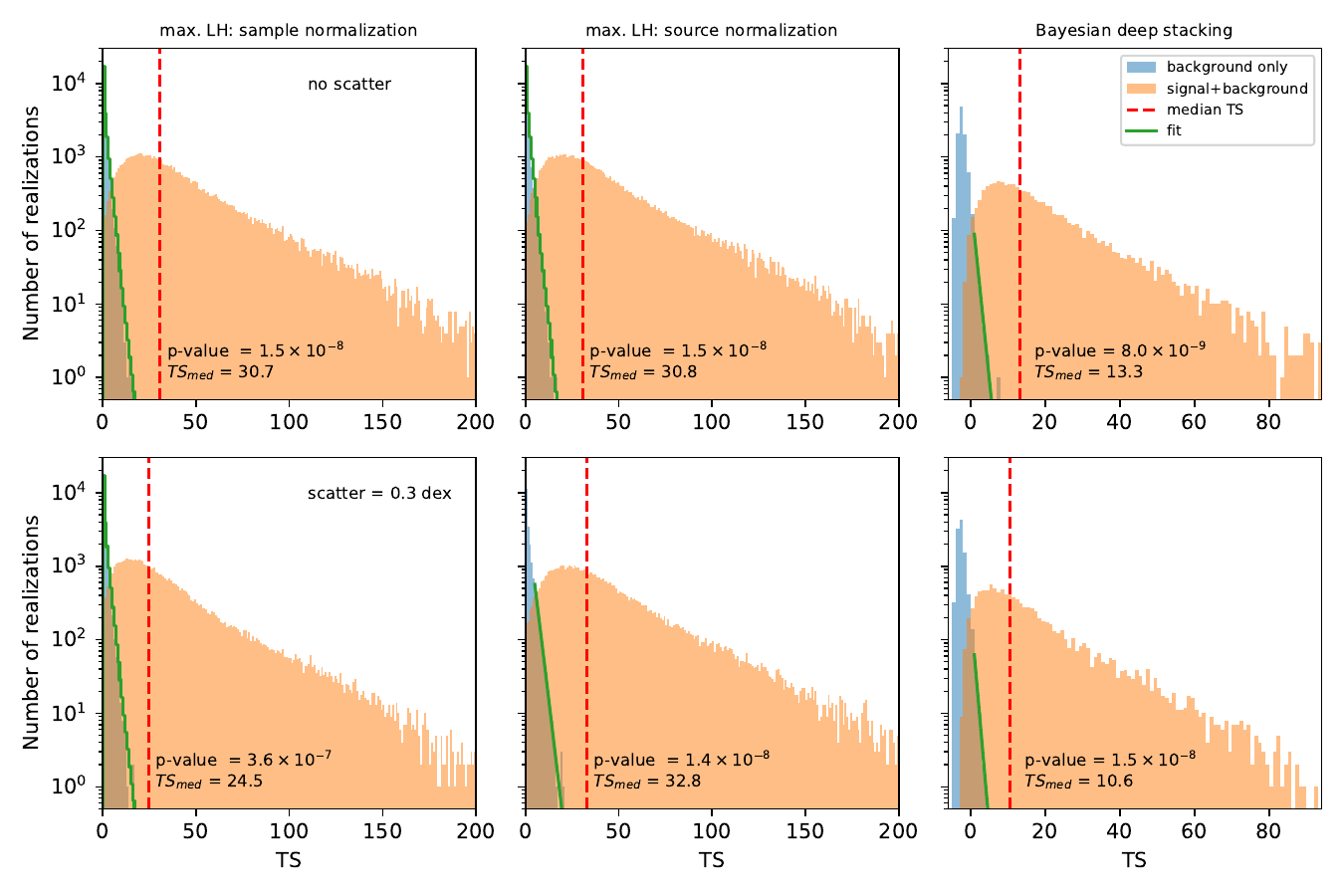}
\includegraphics[width=0.99\textwidth, trim=0 0 0 20, clip]{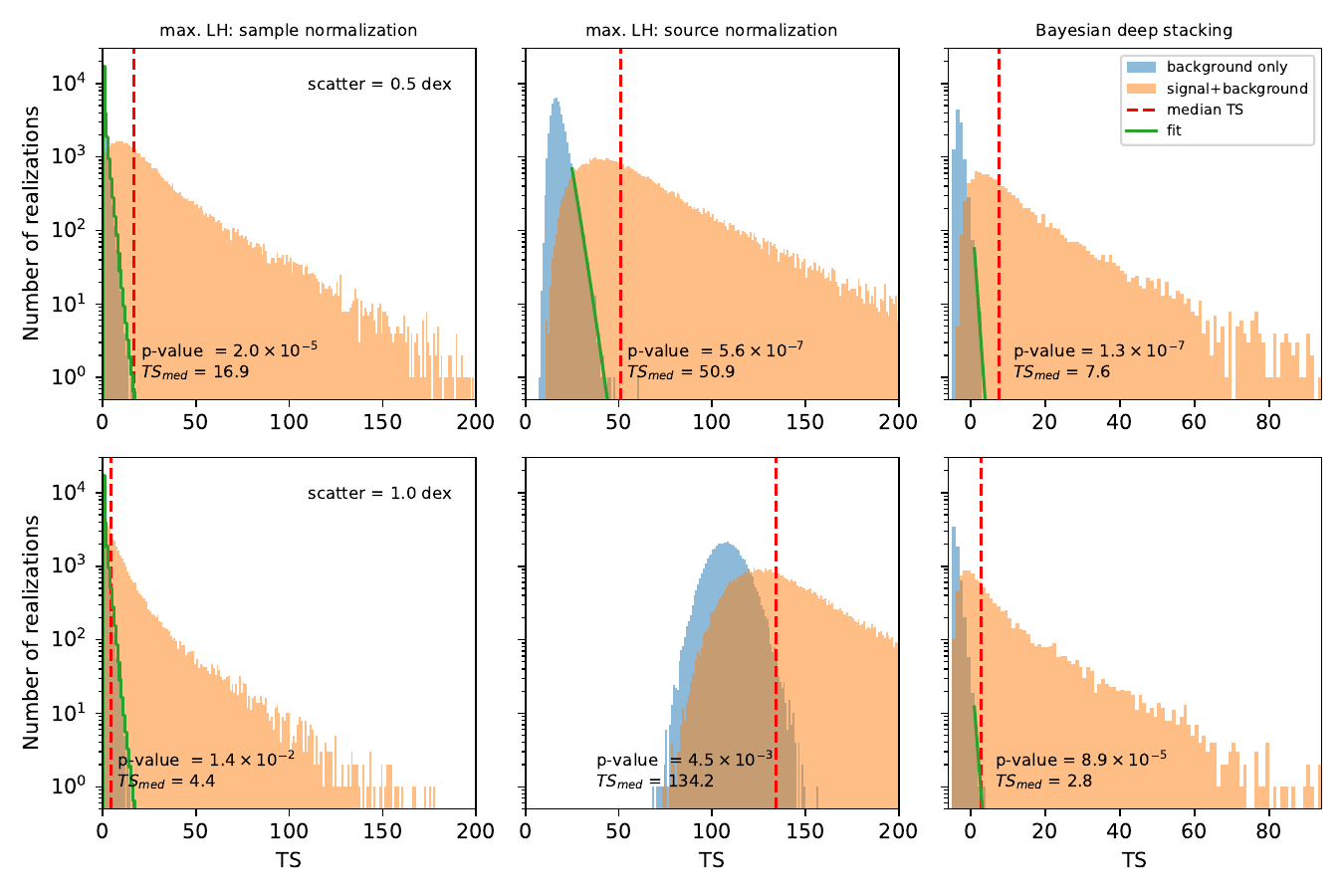}
\vspace{-0.5cm}
    \caption{Test statistic (TS) distributions for different stacking analysis approaches applied to simulated data from the reference neutrino source population described in Section~\ref{sec:simulation}. Rows correspond to different scenarios of scatter between the neutrino luminosity injected into the simulated data and the luminosity assumed in the likelihood analysis: no scatter, 0.3 dex, 0.5 dex, and 1.0 dex, from top to bottom respectively. Columns compare three analysis methodologies: maximum likelihood with sample normalization (left), maximum likelihood with source normalization (center), and Bayesian deep stacking (right). Each panel shows the TS distribution for the background-only (null) hypothesis in blue, and the median TS distribution under the signal-plus-background hypothesis in orange. The p-value indicated in each panel measures the statistical significance of rejecting the null hypothesis at the median TS of the signal distribution ($\mathrm{TS}_{\mathrm{med}}$, indicated by the red dashed line). When necessary, the significance is computed by extrapolating the background-only TS distribution using an appropriate fit (green line, see main text for details).}    
    \label{fig:ts_classic_llh}
\end{figure}

The results of the two maximum likelihood stacking analysis approaches are shown in the left and middle columns of Fig. \ref{fig:ts_classic_llh}. The left column displays results for the standard approach in which we do not take into account flux uncertainty, while the middle column shows results for the source normalization approach. Each row corresponds to a different scatter factor between the injected source flux and the $\hat{n}_{j}$ used in the analysis model ($\sigma_{N}$=0, 0.3, 0.5, 1.0.

For the standard approach, which we refer to as {\it sample normalization}, only one parameter ($\hat{n}_{\rm sig}$) is fitted, and the TS distribution of the background is expected to follow a $\chi^2$-distribution for one degree of freedom. This behavior is confirmed by the simulations, and consequently, we use the $\chi^2$ function to extrapolate the TS distribution. As the scatter increases between simulated signal and the signal model in the analysis, signal-background separation and detection significance deteriorate, as expected. This is particularly visible for scatter factors of 0.5 dex and 1 dex. 

The {\it source normalization} approach is identical to the standard approach for optimal weights, as the priors on the nuissance parameters $\hat{n}_{\rm j}$ approach delta functions\footnote{For numerical stability we use priors with width $\sigma_{N}$=0.01 in this case.} for this scenario. However, in all simulated scenarios where scatter is present between the actual and modeled neutrino flux of the source population, the source normalization approach outperforms the standard approach in terms of detection significance. For example, for a scatter of 0.5 dex, the median detection significance improves from $\approx 4\sigma$ to $\approx 5\sigma$ in the latter case. 

The results from the Bayesian deep-stacking approach presented in this manuscript are finally shown in the right column of Figure  Fig. \ref{fig:ts_classic_llh}. For low scatter factors, the performance of Bayesian deep-stacking is comparable to the source normalization approach, in particular considering the approximate nature of the p-values obtained from extrapolation of the null hypothesis TS distribution (cf., Section \ref{sec:likelihood_analysis}). For larger scatter factors, we find that Bayesian deep-stacking outperforms both frequentist approaches, with the ratio of p-values reaching 50 in favor of Bayesian deep-stacking for a scatter factor of 1.0 dex. In this particular example and scenario, Bayesian deep-stacking is the only approach that can achieve an average detection significance of $>3\sigma$, which is the customary threshold for claiming evidence in high-energy astrophysics. 

It is worth qualitatively investigating how the distributions for the three cases we compare. In the case of no scatter (i.e. source luminosity uncertainty), the three cases give effectively the same background and signal distributions. As we increase scatter, the sample normalization and Bayesian cases behave similarly, with the reduction of the signal test statistic, although this reduction is less for the Bayesian case. Our source normalization case, however behaves differently. Here, both signal and background distributions move to higher test statistics on average. This is due to "overfitting", i.e. the additional degrees of freedom in source luminosities enables better fits that increase the test statistic even if there is no signal.

\section{Conclusion}
\label{sec:conclusion}

We presented a Bayesian framework to carry out deep-stacking analyses for high-energy neutrino observations using source catalogs. We derived the relevant approaches for carrying out both searches and parameter estimation. We also demonstrated the utility of this Bayesian framework over several examples, in comparison to a standard maximum likelihood framework. Our conclusions are the following.
\begin{itemize}
\item A key advantage of the Bayesian framework is that it is straightforward to incorporate all available information about the sources and source models into the analysis. This includes accounting for model uncertainties, e.g., in properly weighing sources during stacking. 
\item We found that the Bayesian framework performs better than a maximum likelihood approach for source parameter estimates (see Section \ref{sec:averageluminosity}) and for search sensitivity. This is due to the incorporation of more information in the Bayesian case. This includes the use of priors, as well as marginalization, which takes into account all available information in posterior distributions. Using only the maximum value of a distribution, as in the maximum likelihood framework, limits the information used and is prone to overfitting, especially over a high-dimensional parameter space. In deep-stacking, we typically encounter many (nuisance) parameters, e.g., the individual flux of each source. Since individual source fluxes are highly uncertain, finding best fits over each of them can lead to overfitting, while treating them as probability densities that are marginalized over remains robust against overfitting. 
\item  We find in particular that Bayesian deep-stacking is superior to frequentist methods in the realistic case that the expected neutrino flux from cataloged sources is uncertain. (see Section \ref{sec:simulation}). Incoporating this uncertainty in a a frequentist framework already helps over standard techniques, but with a Bayesian approach we find potentially large sensitivity gains, reaching $2\sigma$ added sensitivity, or the effective doubling of the signal strenght, for our realistic simulation with $1$\,dex variation (see Fig. \ref{fig:ts_classic_llh}).
\item The Bayesian framework also makes it straightforward to investigate virtually any source properties. While this can be done for the properties of individual sources, it is typically beneficial to instead consider population properties for which the observed data is more informative. We demonstrated the utilization of the Bayesian framework for the reconstruction of such hyperparameter (the average source luminosity).
\end{itemize}

One potential drawback of Bayesian deep-stacking is the increased complexity and computational cost. The latter is especially relevant for cases with large source number densities, in which case the origin of astrophysical neutrinos can be ambiguous as their point spread function overlaps with multiple sources. However, this added complexity is not insurmountable and we anticipate that the added benefits greatly outweigh the cost.

\acknowledgments
We would like to thank Jakob van Santen for valuable feedback on the manuscript. IB is grateful for the support of the National Science Foundation under grant No. PHY-2309024.

\bibliographystyle{JHEP}
\bibliography{my-bib.bib}

\end{document}